\documentclass[%
 reprint,
superscriptaddress,
preprintnumbers,
nofootinbib,
 amsmath,amssymb,
 aps,
prd,
]{revtex4-2}

\usepackage{graphicx}
\usepackage{dcolumn}
\usepackage{bm}
\usepackage{hyperref}

\usepackage{tikz}
\usetikzlibrary{matrix}

\usepackage{aas_macros} 

\newcommand{\figref}[1]{FIG.~\ref{#1}}
\newcommand{\tableref}[1]{TABLE~\ref{#1}}
\newcommand{\appref}[1]{Appendix~\ref{#1}}

\begin{document}

\preprint{CTPU-PTC-25-15}

\title{JFlow: Model-Independent Spherical Jeans Analysis using Equivariant Continuous Normalizing Flows}

\author{Sung Hak Lim}
\email{sunghak.lim@ibs.re.kr}
\affiliation{
    Particle Theory and Cosmology Group,
    Center for Theoretical Physics of the Universe, 
    Institute for Basic Science (IBS),
    Yuseong-gu, Daejeon 34126, Republic of Korea
}
\affiliation{
    Department of Physics and Astronomy, 
    Rutgers,\unpenalty~The State University of New Jersey,
    Piscataway, New Jersey 08854, USA
}
\author{Kohei Hayashi}
\affiliation{
    National Institute of Technology, Sendai College, Sendai, Miyagi 989-3128, Japan
}
\affiliation{
    Astronomical Institute, Tohoku University, Aoba-ku, Sendai 980-8578, Japan
}
\affiliation{
    ICRR, The University of Tokyo, Kashiwa, Chiba 277-8582, Japan
}
\author{Shun'ichi Horigome}
\affiliation{
    Astronomical Institute, Tohoku University, Aoba-ku, Sendai 980-8578, Japan
}
\author{Shigeki Matsumoto}
\affiliation{
    Kavli Institute for the Physics and Mathematics of the Universe (WPI),
    The University of Tokyo Institutes for Advanced Study,
    The University of Tokyo, Kashiwa, Chiba 277-8583, Japan
}
\author{Mihoko M. Nojiri}
\affiliation{
    Theory Center, IPNS, KEK,
    Oho 1-1, Tsukuba, Ibaraki 305-0801, Japan
}
\affiliation{
    Graduate University for Advanced Studies (SOKENDAI),
    Oho 1-1, Tsukuba, Ibaraki 305-0801, Japan
}

\date{\today}

\begin{abstract}
The kinematics of stars in dwarf spheroidal galaxies have been studied to understand the structure of dark matter halos. However, the kinematic information of these stars is often limited to celestial positions and line-of-sight velocities, making full phase space analysis challenging. Conventional methods rely on projected analytic phase space density models with several parameters and infer dark matter halo structures by solving the spherical Jeans equation. In this paper, we introduce an unsupervised machine learning method for solving the spherical Jeans equation in a model-independent way as a first step toward model-independent analysis of dwarf spheroidal galaxies. Using equivariant continuous normalizing flows, we demonstrate that spherically symmetric stellar phase space densities and velocity dispersions can be estimated without model assumptions. As a proof of concept, we apply our method to Gaia challenge datasets for spherical models and measure dark matter mass densities for given velocity anisotropy profiles. Our method can identify halo structures accurately, even with a small number of tracer stars.

\end{abstract}

\maketitle


Understanding the distribution of dark matter around galaxies is crucial for scientific programs trying to identify its fundamental nature.
While galaxy rotation curves clearly demonstrate the presence of gravitational sources beyond visible matter on a galactic scale, the detailed properties of dark matter remain a mystery.
Various experimental efforts aim to understand dark matter, including both direct and indirect detection experiments.
From the particle dark matter perspective, if dark matter consists of weakly interacting particles, these experiments could measure signals from their collisions, annihilations, and decays.

To reveal dark matter's nature, such as its self-interaction cross section or decay rate, understanding its distribution in the space is crucial as these experiments rely on dark matter density.
For example, direct detection experiments \cite{Goodman:1984dc, MarrodanUndagoitia:2015veg, Schumann:2019eaa, Lin:2019uvt} try to measure dark matter collisions using instruments on Earth, so that the signal event rate is proportional to the dark matter density at the solar location.
Indirect detection experiments \cite{Slatyer:2017sev, Hooper:2018kfv} look for the products of dark matter annihilation and decay, and the dark matter density function of the target object determines the signal flux in this case. 
For these experimental searches, precise and accurate estimates of the dark matter density are essential to maximize the discovery potential.

In addition, revealing the dark matter halo structure of galaxies provides crucial insights into the nature of dark matter.
If dark matter exhibits nontrivial interactions, the halo structure may deviate from the conventional Navarro-Frenk-White (NFW) profile \cite{Navarro:1996gj, Navarro:1994hi, Navarro:1995iw}.
For instance, self-interacting dark matter \cite{Tulin:2017ara, Adhikari:2022sbh} predicts a thermal core \cite{Yang:2023jwn}, and wave dark matter \cite{Hui:2021tkt} predicts a soliton core and oscillating eigenmodes \cite{Yavetz:2021pbc}.
To cover the possibility of non-trivial halo shapes and to estimate shape uncertainties, which are not fully accounted for in parametric fitting with conventional stellar and dark matter density profiles, we need a more general approach with fewer assumptions.

For these purposes, we propose a new unsupervised machine learning technique ``\texttt{JFlow}" for analyzing dark matter halos in approximately spherical stellar systems, particularly for studying dwarf spheroidal galaxies.
Dwarf spheroidal galaxies (dSphs) are small and faint galaxies; many of them are found as satellites of the Milky Way.
These galaxies consist of old stellar populations with little intergalactic dust, making all sufficiently luminous stars clearly visible.
Moreover, stars in dSphs contribute only a tiny fraction of the total mass; therefore, the gravitational fields are dominated by their dark matter halos, and stars act as ideal tracers of the gravitational potential. 
Additionally, dSphs exhibit minimal baryonic activity in their cores, allowing dark matter signals from their galactic centers to be detected with less background contamination.
Given their high dark matter content and role as clean signal sources, dSphs are ideal targets for indirect detection experiments, and their halo structure can be inferred through the kinematics of their member stars.

Despite the advantages of dSphs as dark matter laboratories, assumption-free inference of the dark matter halo is challenging due to data limitations.
For individual stars in these systems, we can only measure their projected positions on the sky and line-of-sight velocities precisely, as parallaxes and proper motions of distant stars are hard to measure.
The full stellar phase-space information remains inaccessible at the raw data level.
Due to this limitation, it is difficult to directly estimate the gravitational field and matter density by solving statistical equations of motion for the stellar phase-space density function, such as the collisionless Boltzmann equation.

In recent years, unsupervised machine learning methods based on normalizing flows \cite{kobyzev2020normalizing} have demonstrated success in free-form phase-space density and mass density estimation.
These methods have been validated on both idealized systems \cite{2020arXiv201104673G, GanDSPH, 10.1093/mnras/stab2049, 2022MNRAS.511.1609N, 2023ApJ...942...26G} and real applications, including dark matter density estimation in the solar neighborhood using the $N$-body simulated galaxy \texttt{h277} \cite{Buckley:2022tjy, h277, 2012ApJ...761...71Z, 2012ApJ...758L..23L} and stellar kinematics from the Gaia DR3 dataset presenting stars in Milky Way \cite{Lim:2023lss, Putney:2024tfq}.
However, applying these methods to dSphs requires reconstructing full 6D phase-space information from limited kinematic data.

To overcome these observational limitations, the traditional approach assumes spherical symmetry and applies tomographic deprojections.
For approximately spherical systems like dSphs, the Abel transformation is the simplest method for density deprojection assuming spherical symmetry.
The Abel transformation relates the 3D stellar number density $n(r)$ at the distance $r$ from the galactic center and the 2D projected stellar surface density, $I(R)$, evaluated at the projected distance $R$ to the orthogonal plane of the line-of-sight direction. 
The inverse transformation from $I(R)$ to $n(r)$ is given as follows:
\begin{equation}
    n(r) = - \frac{1}{\pi} \int_r^\infty \frac{dR}{\sqrt{R^2-r^2}} \frac{dI}{dR}
    \label{eqn:abel_deprojection}
\end{equation}
Using this transformation, we can estimate the three-dimensional stellar number density $n(r)$ from the observed projected density $I(R)$. 
For the velocity distribution deprojection, we refer to a review in \cite{Ullio_2016}.

Using the spherically symmetric deprojections, the conventional method for estimating dark matter halo structure in dSphs is based on solving the spherical Jeans equation.
Assuming that the stellar system is in kinetic equilibrium, the spherical Jeans equation relates the stellar phase space density and dark matter density as follows:
\begin{equation}
\frac{d}{d r} n \overline{v_r^2}+ \frac{2 \beta}{r} n  \overline{v_r^2} = - n \frac{d \Phi}{dr}
, \quad
\beta = 1 - \frac{\overline{v_\theta^2}+\overline{v_\phi^2}}{2\overline{v_r^2}},
\end{equation}
where $\overline{v_r^2}(r)$, $\overline{v_\theta^2}(r)$, $\overline{v_\phi^2}(r)$ are the second moments of radial, polar, azimuthal velocities at the distance $r$, resplectively; $\beta(r)$ is the velocity anisotropy function characterizing the ratio between radial and tangential velocity dispersions; and $\Phi(r)$ is the gravitational potential.

For dSphs where the dark matter halo dominates the gravitational field, the acceleration $-d \Phi(r) / dr $ is fully determined by the enclosed dark matter mass $M(r)$, i.e., $d \Phi(r) / dr = G M(r) / r^2$.
The closed-form expression of $M(r)$ from the spherical Jeans equation is as follows:
\begin{equation}
    M(r) = -\frac{r \overline{v_r^2}}{G} \left( 
        \frac{d \ln(n \overline{v_r^2})}{d \ln r}
        + 2 \beta
    \right).
    \label{eqn:jeans_mass}
\end{equation}
This expression relates the dark matter density of dSphs to their stellar phase space density.
Note that conventional parametric fitting methods use predefined stellar density, dark matter halo, and velocity anisotropy profiles; compute projected stellar density and velocity dispersions using the Abel transformations and spherical Jeans equation; and compare these predictions with data.\footnote{Examples of parametric analysis spans various levels of complexity from simple functional models like \cite{Hayashi:2016kcy, Ichikawa:2016nbi, Ichikawa:2017rph, Horigome:2020kyj} to complex but expressive models like \texttt{JEAnS} modeling using B-splines \cite{2017MNRAS.470.2034D, 2019MNRAS.482.3356D} and \texttt{gravsphere} using Plummer mixture density models for stellar number density and piecewise function models for dark matter density \cite{2017MNRAS.471.4541R, 2018MNRAS.481..860R, 2020MNRAS.498..144G, 2021MNRAS.505.5686C}.}

The Jeans equation in the form of \eqref{eqn:jeans_mass} tells us that the three functions, $n(r)$, $\overline{v_r^2}(r)$, and $\beta(r)$, completely determine the spherical dark matter mass distribution.
In the following, we propose the \texttt{JFlow} method, which uses neural networks to directly model the stellar number density $n(r)$ and radial velocity dispersion $\overline{v_r^2}(r)$ and estimate the dark matter distribution via \eqref{eqn:jeans_mass}.
Note that the anisotropy function cannot be estimated from the line-of-sight velocity data alone because knowing only line-of-sight velocity dispersion is insufficient to determine multiple velocity dispersion components.
This ambiguity allows multiple enclosed mass solutions in Jeans analysis for dSphs, known as mass-anisotropy degeneracy.
Since this letter mainly aims to generalize the classical spherical Jeans analysis for the proof-of-concept, we will provide the anisotropy function $\beta(r)$ as an input to resolve the degeneracy.

The \texttt{JFlow} model starts from the stellar number density estimation using a neural density estimator called normalizing flows \cite{kobyzev2020normalizing}.
This neural network is trained to learn a transformation from a simple base distribution\footnote{In the context of normalizing flows, simple distributions refer to those that are easy to sample from and evaluate density functions.} to the target data distribution to be estimated.
Specifically, if $T$ is a neural network with a set of parameters, $\theta$, the following transformation relates a point $\vec{r}_0$ sampled from the base distribution to the associated data point $\vec{r}_1$:
\begin{equation}
    \vec{r}_1 = T(\vec{r}_0; \theta)
\end{equation}
The likelihood/probability density function can then be evaluated using the change of variable formula: 
\begin{equation}
    p(\vec{r}_1) = p(\vec{r}_0) \left|\frac{ \partial T(\vec{r}_0; \theta)}{ \partial \vec{r}_0} \right|
\end{equation}
We choose the standard normal distribution as the base distribution due to its simplicity, i.e., $p(\vec{r}_0)$ is the 3D standard Gaussian probability density function.
After finishing the training, the learned target distribution $p(\vec{r}_1)$ will regress the number density $n(r)$.

We use continuous normalizing flows \cite{NEURIPS2018_69386f6b, grathwohl2018scalable} to make this architecture only learn spherically symmetric densities.
This neural network models an infinitesimal coordinate transformation described by the following ordinary differential equation (ODE),
\begin{equation}
    \frac{d \vec{r}_t}{dt} = \vec{F}(\vec{r}_t, t; \theta),
\end{equation}
where $\vec{r}_t$ is the coordinate at time $t$, and $\vec{F}$ is a neural network.
The time $t=0$ represents the data from the base distribution, and $t=1$ refers to the data distribution.
The transformation $T$ is the solution of this ODE.

As the standard Gaussian distribution is already spherically symmetric, the learned density will be spherically symmetric if the transformation $T$ is equivariant under rotation.
In this letter, we consider only the radial directional transformation as follows,
\begin{equation}
    \frac{d \vec{r}_t}{dt} = \hat{r} F_r(|\vec{r}_t|, t; \theta),
\end{equation}
where $\hat{r}$ is the radial unit vector.
Then, polar and azimuthal coordinates remain unchanged, preserving spherical symmetry.
This realization of equivariant continuous normalizing flows \cite{pmlr-v119-kohler20a, 2019arXiv191000753K, Kanwar:2020xzo} can model any spherically symmetric density.

Since the stellar distributions of dSphs often exhibit a simple flat core \cite{1988AJ.....95.1706E, 1988AJ.....96.1352E, 10.1093/mnras/277.4.1354} like the Plummer model \cite{1911MNRAS..71..460P}, we constrain our continuous normalizing flows to cover only cored profiles by suppressing the transformation below a transition length scale $r_s$.
For example, we consider the following ODE model:
\begin{equation}
\frac{d \vec{r}_t}{dt} = \hat{r} \tanh \frac{|\vec{r}_t|}{r_s} \cdot F_r(|\vec{r}_t|, t; \theta).
\label{eqn:cnf_core}
\end{equation}
With this $\tanh$-suppressed transformation, the density around the core remains constant as the base Gaussian distribution also has a flat inner core. 
We denote the corresponding transformation of this ODE as $T_{\mathrm{core}}$.

Although dSphs often exhibit cored stellar density profiles, spherically symmetric galactic systems in general may exhibit other non-trivial behaviors like power-law cusps \cite{10.1093/mnras/202.4.995, 1990ApJ...356..359H, 10.1093/mnras/265.1.250} or Sérsic profiles with varying logarithmic slopes $d\log n(r) / d\log r$ \cite{1963BAAA....6...41S, 1948AnAp...11..247D, 2003AJ....125.2951G}.
To consider cuspy profiles with non-vanishing logarithmic slope at the core, we additionally apply a power-law transformation to the radial component of the cored density model.
The transformation with exponent $c+1$ and its Jacobian determinant is as follows,
\begin{equation}
    T_{\mathrm{pow}}: |\vec{r}| 
    \rightarrow 
    |\vec{r}|^{c+1}, \quad 
    \left| \frac{\partial T_{\mathrm{pow}}(\vec{r})}{\partial \vec{r}} \right|
    = (c+1)  |\vec{r}|^{-\frac{D c}{c+1}},
    \label{eqn:trans_pow}
\end{equation}
where $D=3$ is the spatial dimension. 
If $c=0$, the transformation is the identity, and the modeled distribution's core remains flat.
Otherwise, the Jacobian determinant of the composite transformation $T_{\mathrm{pow}} \circ T_{\mathrm{core}}$ will introduce the power-law scaling cusp around the core.
We use the resulting combined mapping ${T_\mathrm{cusp} = T_{\mathrm{pow}} \circ T_{\mathrm{core}}}$ as the transformation in our normalizing flows to model both cored and cuspy stellar density profiles.

Our 3D stellar density model evaluated on Cartesian coordinates may seem less direct than performing 1D density estimation on only the radial component, but this setup provides better numerical stability, especially near the core.
In dSph analysis, precise inner density estimation is crucial as it plays an important role in both indirect dark matter detection and dark matter model inference through core shape analysis.
The spherical coordinate system has a coordinate singularity at the origin, which is the coordinate boundary.
Density estimation near the singularity becomes less reliable due to discontinuity issues when using $|\vec{r}|$ and data sparsity when using $\log |\vec{r}|$, significantly affecting the precision of free-form density estimation near the origin.
Direct 3D density estimation has better precision as the 3D Cartesian coordinate system does not have coordinate singularities and the origin is not at the boundary; the prediction of inner profiles becomes more stable with this setup.

For the velocity dispersion modeling, we use a conditional linear flow for the variance estimation.
This flow models Gaussian distributions conditioned on positions by applying a position-dependent linear transformation to the standard normal base distribution.
Since the maximum likelihood estimate of a Gaussian's variance parameter gives the sample variance, this parametrization naturally provides a method for velocity dispersion estimation.
We implement this using the following position-dependent velocity transformation in spherical coordinates:
\begin{eqnarray}
    \vec{v} & \rightarrow & L(r; \theta) \cdot \vec{v},
    \label{eqn:vel_dispersion}
    \\
    \nonumber
    L(r; \theta) & = & \left(\begin{matrix}
        \sqrt{\overline{v^2_r}}(r; \theta) & 0 & 0 \\
        0 & \sqrt{\overline{v^2_\theta}}(r; \theta) & 0 \\
        0 & 0 & \sqrt{\overline{v^2_\phi}}(r; \theta) 
    \end{matrix}\right),
\end{eqnarray}
where the velocity dispersions in each direction are modeled by neural networks as functions of radius.
We use a multilayer perceptron to model the decimal logarithm of radial velocity dispersion, $\log_{10} \overline{v^2_r}(r; \theta)$, and the tangential velocity dispersions are then determined using the provided anisotropy profile:
\begin{equation}
    \overline{v^2_\theta}(r; \theta) = \overline{v^2_\phi}(r; \theta) = \overline{v^2_r}(r; \theta) \cdot ( 1- \beta(r)).
\end{equation}


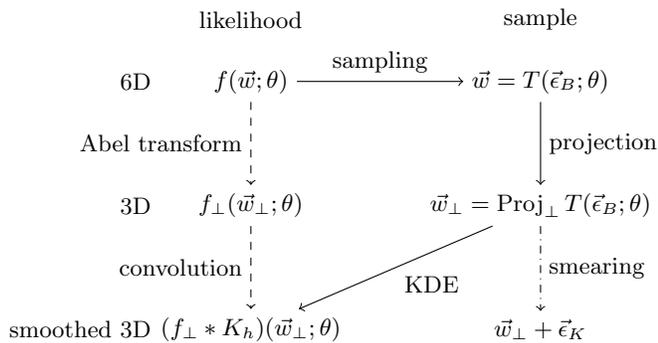
\begin{figure}
\begin{center}
\begin{tikzpicture}[baseline={([yshift=-.5ex]current bounding box.center)},vertex/.style={anchor=base,
    circle,fill=black!25,minimum size=18pt,inner sep=2pt},scale=1.1,yscale=-1]
\node [draw=none] (likelihood_label)   at (0.0,-0.75) {likelihood};
\node [draw=none] (sample_label)       at (3.5,-0.75) {sample};
\node [draw=none, anchor=east] (6D_label)       at (-1.1,0.0) {6D};
\node [draw=none, anchor=east] (3D_label)       at (-1.1,1.5) {3D};
\node [draw=none, anchor=east] (3Dconv_label)   at (-1.1,3.0) {smoothed 3D};
\node [draw=none] (6D_prob)   at (0,0) {$f(\vec{w}; \theta)$};
\node [draw=none] (6D_sample) at (3.5,0) {$\vec{w} = T(\vec{\epsilon}_B;\theta)$};
\node [draw=none] (3D_prob)   at (0,1.5) {$f_\perp(\vec{w}_\perp; \theta)$};
\node [draw=none] (3D_sample) at (3.5,1.5) {$\vec{w}_\perp = \mathrm{Proj}_\perp \, T(\vec{\epsilon}_B;\theta)$};
\node [draw=none] (3Dconv_prob)   at (0,3.0) {$(f_\perp*K_h)(\vec{w}_\perp; \theta)$};
\node [draw=none] (3Dconv_sample) at (3.5,3.0) {$\vec{w}_\perp + \vec{\epsilon}_K$};
\draw [->, dashed] (6D_prob) -- (3D_prob)
    node [midway,left,align=center] {Abel transform};
\draw [->, dashed] (3D_prob) -- (3Dconv_prob)
    node [midway,left,align=center] {convolution};
\draw [->] (6D_prob) -- (6D_sample)
    node [midway,above,align=center] {sampling};
\draw [->] (6D_sample) -- (3D_sample)
    node [midway,right,align=center] {projection};
\draw [->] (3D_sample) -- (3Dconv_prob)
    node [midway,below right,align=center] {KDE};
\draw [->, dash dot] (3D_sample) -- (3Dconv_sample)
    node [midway,right,align=center] {smearing};
\end{tikzpicture}
\end{center}
\caption{\label{fig:diagram}
Schematic diagram of different approaches for modeling likelihood functions in smoothed 3D observable space.  
Dashed arrows show the conventional path using the Abel transformation. 
Solid arrows show our approach using sampling from normalizing flows and kernel density estimation (KDE) subsequently. 
The dot-dashed arrow indicates kernel smoothing applied to training samples to match the smoothing of the likelihood function.
}
\end{figure}

Having described our model for a 6D phase space density for Jeans analysis, we now explain how to train the networks with limited 3D kinematic information $\vec{w}_\perp$ derived from positions on the sky and line-of-sight velocities.
While one might consider applying Abel transformations to the phase space density to construct likelihood functions for the training (the direction illustrated as dashed arrows in \figref{fig:diagram}), this approach is challenging due to the complex parametrization of neural networks.
Another approach is to use a generative adversarial network \cite{NIPS2014_5ca3e9b1} for the 3D space, as the samples generated by normalizing flows are differentiable with respect to the network parameters \cite{GanDSPH}.
However, generative adversarial networks require training an additional classifier and introduce training complexity, motivating us to seek a simpler strategy.

As the dimensionality of the dataset is low, we directly reconstruct the likelihood function $f_\perp$ in 3D observable space using kernel density estimation (KDE) on the projected samples generated by the normalizing flows.
We denote the KDE-estimated likelihood function as $\hat{f}_\perp * K_h$, where $\hat{f}_\perp$ is the likelihood function estimator, and $K_h$ is scaled Gaussian kernel with bandwidth parameter $h$.
We illustrate this procedure for reconstructing the projected likelihood function with solid arrows in \figref{fig:diagram}.
Since this likelihood function model is differentiable, we can use it to train the network.

The kernel density estimation additionally introduces a smoothing effect that must be incorporated into both the training dataset and loss function for proper network training.
Since we have a kernel-smoothed likelihood function, the training dataset should be smeared by the same kernel, as illustrated by the dot-dashed arrow in \figref{fig:diagram}. 
This ensures both likelihood functions and training samples are in the same smoothed 3D observable space.
The resulting asymptotic form of the negative log-likelihood loss function $\mathcal{L}(\theta)$ for training the network parameters is:
\begin{equation}
    \mathcal{L}(\theta) = - \int d\vec{w}_\perp \, (f_{\perp} * K_h) (\vec{w}_\perp) \log (\hat{f}_{\perp} * K_h) (\vec{w}_\perp; \theta).
    \label{eqn:loss}
\end{equation}
The convoluted distribution $f_{\perp} * K_h$ represents the smoothed empirical distribution that will be evaluated using smeared training datasets.

This loss function is equivalent to the Kullback-Leibler divergence between the smoothed densities $f_{\perp} * K_h$ and $\hat{f}_{\perp} * K_h$.
Therefore, the optimal solution of the training in the asymptotic limit is $f_{\perp} * K_h = \hat{f}_{\perp} * K_h$, and $\hat{f}_{\perp}$ will converge to $f_{\perp}$ in the case of  Gaussian kernels.
We discuss evaluating this loss function with finite samples and details of the training setup in \appref{app:nn_details}.

For a proof-of-concept of our machine learning technique, we demonstrate the method using the spherical test suite of the Gaia Challenge dataset \cite{gaiaChallengeSph, 2011ApJ...742...20W}. 
This dataset provides mock spherical stellar systems with known stellar and dark matter halo profiles, making it suitable for validating our method's ability to recover true mass density profiles. 
In this letter, we present results for the mock dataset \texttt{NonplumCoreIso} \cite{gaiaChallengeSph}, which has cuspy stellar density, cored dark matter halo, and isotropic velocity dispersions.
To test the applicability of our algorithm for faint dSphs, we choose the dataset with 1,000 samples.
The details of this mock dataset are described in \appref{app:dataset}.

\begin{figure*}[t]
    \centering
    \includegraphics[width=0.245\linewidth]{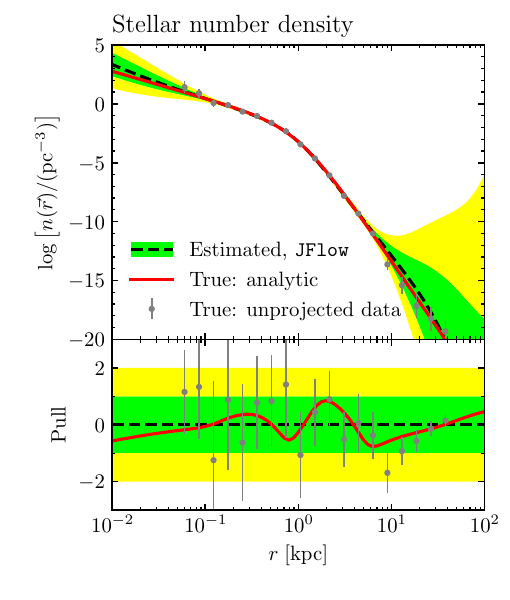}
    \includegraphics[width=0.245\linewidth]{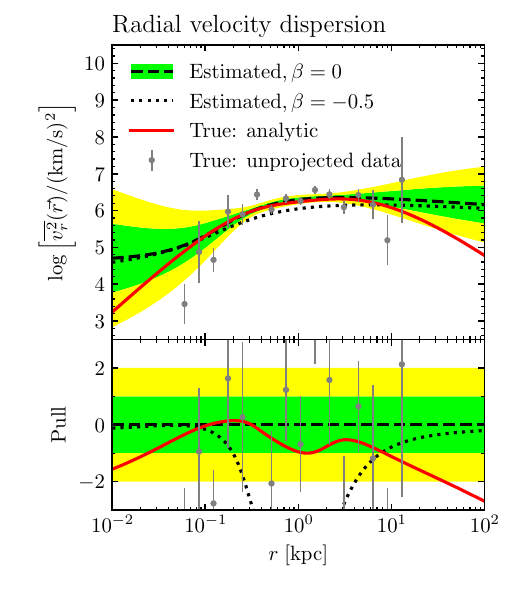}
    \includegraphics[width=0.245\linewidth]{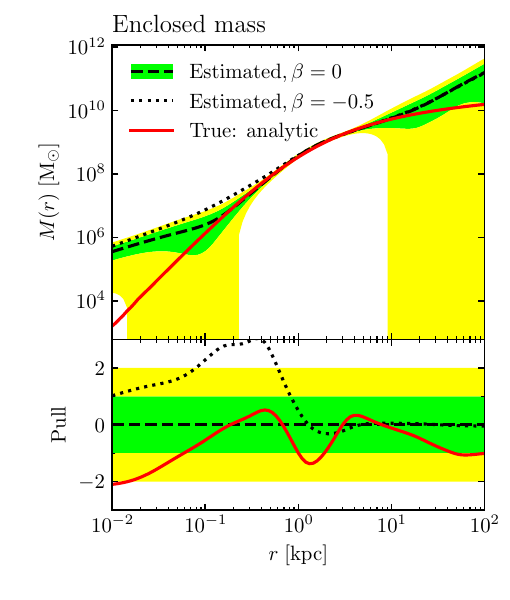}
    \includegraphics[width=0.245\linewidth]{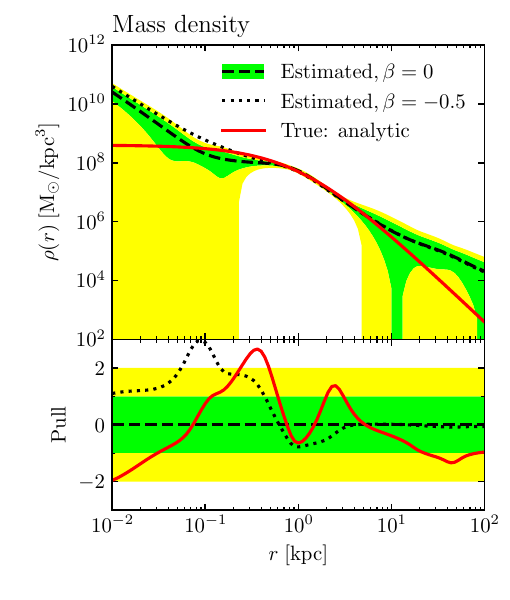}
    \caption{
        \texttt{JFlow} predictions for dSph parameters of \texttt{NonplumCoreIso} dataset.
        Panels show (from left to right): stellar number density, radial velocity dispersion, enclosed mass, and total mass density as functions of galactocentric radius. 
        Black dashed lines indicate \texttt{JFlow}'s best-fit values using the correct velocity anisotropy function $\beta=0$. 
        Green and yellow bands show $1\sigma$ and $2\sigma$ confidence intervals estimated by the bootstrap method. 
        Black dotted lines are \texttt{JFlow}'s best-fit values using an incorrect velocity anisotropy function $\beta=-0.5$.
        Stellar number density results for $\beta=-0.5$ are omitted as they are independent of the velocity anisotropy $\beta$.
        Red lines represent true values from the simulation.
        For stellar density and radial velocity dispersion, gray error bars show $1\sigma$ uncertainties computed from histograms of full 6D phase-space data. 
        Lower panels show pull values, i.e., residuals from \texttt{JFlow} estimations normalized by \texttt{JFlow}'s $1\sigma$ uncertainties.   
    }
    \label{fig:jflow_res}
\end{figure*}

\figref{fig:jflow_res} shows the predictions from our \texttt{JFlow} algorithm. 
The green and yellow bands represent $1\sigma$ and $2\sigma$ confidence intervals estimated by the bootstrap method. 
To obtain these intervals, we resample the dataset 20 times, apply the \texttt{JFlow} algorithm to each resampled dataset, and use the standard deviation of the predictions as the $1\sigma$ uncertainty.
These uncertainties cover both statistical uncertainties and systematic variations from neural network initialization. 
Our estimations generally agree with the true values within $2\sigma$.
Note that results at galactocentric radii $r < 0.4$ kpc and $r > 20$ kpc are less reliable due to the smaller number of samples in these regions. 
\texttt{JFlow} assigns larger uncertainties in these regions, as expected for free-form models in extrapolated regions.

The largest deviation from the true value appears in the mass density estimation.
The deviation of $2.5\sigma$ at 0.4 kpc is from the overestimated curvature of the radial velocity distribution near the boundary of available training data.
We expect this deviation would be reduced by regularizing the radial velocity dispersion model's behavior at small $r$ (similar to our approach for cored and cuspy stellar number density models) or by using larger samples.
Introducing systematic smoothing at small length scales, as in \cite{Buckley:2022tjy, Lim:2023lss}, may further improve these short-distance density estimates.
We leave systematically incorporating such smoothing methods with \texttt{JFlow} for future studies.

Black dotted lines in \figref{fig:jflow_res} show \texttt{JFlow} estimates using incorrect anisotropy profiles to test the sensitivity to anisotropy assumptions.
We use a slightly tangentially biased anisotropy, $\beta=-0.5$. 
As this assumption is incompatible with the true anisotropy $\beta=0$, we observe a significant deviation in radial velocity dispersion estimation, which propagates to the enclosed mass and total mass density estimates.
Nevertheless, the enclosed mass estimation at $r=1$ kpc remains consistent with the true value, as this corresponds to the Wolf radius \cite{wolf2010accurate} where the enclosed mass estimate from Jeans analysis becomes independent of $\beta$.
The mass density estimate at $r=1$ kpc is also roughly consistent with the true value, but this is just because the enclosed mass estimate remains close to the true value in the vicinity of the Wolf radius. The results may change for different anisotropy profiles.

We additionally compare these results with those of conventional parametric inferences at \appref{app:map}.
We may also compare our method with simulation-based inferences \cite{Nguyen:2022ldb, 2025arXiv250303812N} that fully utilize all the information beyond number density and velocity dispersions, but the comparison is beyond the scope of this letter.

In summary, we have introduced an unsupervised machine learning method \texttt{JFlow} extending classical spherical Jeans analysis. 
The conventional spherical Jeans analysis for revealing dark matter halos in systems like dwarf spheroidal galaxies assumes specific stellar and dark matter density profiles, requiring model extensions when considering halos with non-trivial dark matter interactions or uncertainties on the profile choice itself.
Our approach addresses these limitations through normalizing flows for free-form density estimation.
\texttt{JFlow} is fully model-independent, enabling analysis of various physics scenarios with distinct stellar and dark matter density profiles.
To train \texttt{JFlow} using limited kinematic data of dwarf spheroidal galaxies, we developed a likelihood-free training method based on generative modeling and kernel density estimation.
This training setup is highly adaptable to complex situations, making it valuable for future extensions to non-spherical systems and considering observational effects where likelihood projection is challenging.

While we have presented a proof-of-concept example, the flexibility of our approach suggests natural extensions to non-spherical systems, as well as extensions beyond first-order Jeans analysis, such as higher-order Jeans analysis \cite{2020MNRAS.498..144G} and multi-component analysis \cite{2011ApJ...742...20W} for alleviating mass-anisotropy degeneracy.
We anticipate that our \texttt{JFlow} method will open new possibilities for machine learning-based analysis of distant galaxy dynamics.

\emph{Acknowledgement} --
This work was supported by IBS under the project code, IBS-R018-D1.
The work of SHL was also partly supported by the DOE under Award Number DOE-SC0010008.
This research used resources of the National Energy Research Scientific Computing Center, a DOE Office of Science User Facility supported by the Office of Science of the U.S. Department of Energy under Contract No. DE-AC02-05CH11231 using NERSC award HEP-ERCAP0027491.
This work was also performed in part at Aspen Center for Physics, which is supported by National Science Foundation grant PHY-2210452. 
This work is supported by Grant-in-Aid for Scientific Research from the Ministry of Education, Culture, Sports, Science, and Technology (MEXT), Japan, grant numbers JP24K00669 and JP25H01553.
This work is also supported by Grant-in-Aid for Transformative Research Area (A) 22H05113
and Grant-in-Aid for Scientific Research(C) JSPS KAKENHI Grant Number 22K03629.
SH is also supported by a Grant-in-Aid for Early-Career Scientists JSPS KAKENHI Grant Number 23K13098 and Grant-in-Aid for JSPS Fellows JSPS KAKENHI Grant Number 25KJ0017.
SM is also supported by a Grant-in-Aid for Scientific Research from the MEXT, Japan (24H00244, 24H02244) and by the World Premier International Research Center Initiative (WPI), MEXT, Japan (Kavli IPMU).
The authors acknowledge the Office of Advanced Research Computing (OARC) at Rutgers, The State University of New Jersey for providing access to the Amarel cluster and associated research computing resources that have contributed to the results reported here (URL: \url{https://oarc.rutgers.edu}).

\bibliography{jflow}

\appendix

\section{\texttt{NonplumCoreIso} Datasets}
\label{app:dataset}
Here we describe the simulation parameters of the \texttt{NonplumCoreIso} dataset \cite{gaiaChallengeSph}.
This mock dataset is generated using generalized Hernquist density profiles \cite{1990ApJ...356..359H, 1996MNRAS.278..488Z} for both stellar number density $n(r)$ and dark matter density $\rho_\mathrm{DM}(r)$ at galactocentric radius $r$.
The profiles are characterized by inner and outer scaling exponents $-\gamma$ and $-\beta$:
\begin{equation}
    n(r) \propto \left(\frac{r}{r_*}\right)^{-\gamma_*} \left[ 1+ \left(\frac{r}{r_*}\right)^{\alpha_*}\right]^{\frac{\gamma_*-\beta_*}{\alpha_*}},
    \label{eqn:model:stellar}
\end{equation}
\begin{equation}
    \rho_{\mathrm{DM}} (r) = \rho_0 \left(\frac{r}{r_{\mathrm{DM}}}\right)^{-\gamma_{\mathrm{DM}}} \left[ 1+ \left(\frac{r}{r_{\mathrm{DM}}}\right)^{\alpha_{\mathrm{DM}}}\right]^{\frac{\gamma_{\mathrm{DM}}-\beta_{\mathrm{DM}}}{\alpha_{\mathrm{DM}}}}.
    \label{eqn:model:dark_matter}
\end{equation}
Here $r_*$ and $r_\mathrm{DM}$ are the characteristic radii where the density profiles transition from inner slope $-\gamma$ to outer slope $-\beta$, with $\alpha$ controlling the sharpness of this transition.
$\rho_0$ is the normalization parameter for the dark matter density profile.
The parameter values used in the simulation are listed in \tableref{tab:prior}.

\begin{table}[]
    \caption{
    Prior ranges and true values of the generalized Hernquist profile parameters for the \texttt{NonplumCoreIso} dataset. 
    The second column shows uniform prior ranges used in the maximum a posteriori estimation, and the third column shows the true parameter values from the simulation.
    }
    \centering
    \begin{ruledtabular}
    \begin{tabular}{ccc}
         Parameter & Support intervals & True value\\
         \hline
         $\alpha_*$ & [ $\phantom{-}0.5$ , \phantom{1}3 ] & 2\\
         $\beta_*$ & [ $\phantom{-}3\phantom{.0}$ , 10 ] & 5\\
         $\gamma_*$ & [ $\phantom{-}0\phantom{.0}$ , \phantom{1}3 ] & 1\\
         $\log_{10} \left[r_* / \mathrm{kpc} \right]$ & [ $-2\phantom{.0}$ , \phantom{1}2 ] & $\log_{10} \left[ 1 \right] = 0$\\
         \hline
         $\alpha_{\mathrm{DM}}$ & [ $\phantom{-}0.5$ , \phantom{1}3 ] & 1\\
         $\beta_{\mathrm{DM}}$ & [ $\phantom{-}3\phantom{.0}$ , 10 ] & 3 \\
         $\gamma_{\mathrm{DM}}$ & [ $\phantom{-}0\phantom{.0}$ , \phantom{1}3 ] & 0\\
         $\log_{10} \left[ r_{\mathrm{DM}} / \mathrm{kpc} \right]$ & [ $-2\phantom{.0}$ , \phantom{1}2 ] & $\log_{10} \left[ 1 \right] = 0$ \\
         $\log_{10} \left[ \rho_{0} / (\mathrm{M_\odot / kpc^3})\right]$  & [ $-3\phantom{.0}$ , \phantom{1}3 ] & $\log_{10} \left[ 0.4 \right] \approx -0.398$\\
    \end{tabular}
    \end{ruledtabular}
    \label{tab:prior}
\end{table}

\section{Neural Network Model and Training Details}
\label{app:nn_details}

Our \texttt{JFlow} algorithm uses two multilayer perceptrons (MLPs) to model $F_r(|\vec{r}_t|, t)$ in \eqref{eqn:cnf_core} and $\log_{10} \sqrt{\overline{v_r^2}}(r)$ in \eqref{eqn:vel_dispersion}, trained using standardized datasets.
The first MLP models $F_r(|\vec{r}_t|, t)$ with 5 hidden layers of width 32, taking $(|\vec{r}_t|, t)$ as input.
The second MLP models $\log_{10} \sqrt{\overline{v_r^2}}(r)$ with 4 hidden layers of width 32, taking $\log_{10}|\vec{r}|$ as input.
These MLPs contain 4,353 and 3,265 parameters, respectively.
Both networks use GELU activation functions \cite{hendrycks2016gaussian}.
Given the limited sample size of 1,000, we avoid too high over-parametrization, and the setup is sufficient, as demonstrated in the main text.
The implementation uses the \texttt{PyTorch} library.

The loss function in \eqref{eqn:loss} is evaluated as follows. 
We first reserve 50\% of the dataset for training and use the remaining data for validation.
The integral is evaluated using a (quasi) Monte-Carlo method, where the integral over smoothed density is approximated by the following summation:
\begin{equation}
    \int d\vec{w}_\perp \, (f * K_h) (\vec{w}_\perp)
    \rightarrow
    \frac{1}{N N_K} \sum_{a=1}^N \sum_{b=1}^{N_K},
\end{equation}
where we sum over $N$ training/validation samples $\vec{w}_\perp^{(a)}$ and $N_K$ kernel samples $\vec{\epsilon}^{(b)}$.
The kernel samples are quasi-randomly generated: for the 1D problem (number density fitting), we use inverse transform sampling with 50 linearly spaced points on $[0,1]$ excluding boundaries; for higher-dimensional problems (velocity dispersion fitting), we apply inverse transform sampling on each axis using the first 50 entries of Halton sequences.
The total number of smeared samples is $N \times N_K = 500 \times 50 = 25,000$.
For training, these samples are randomly split into 5 mini-batches of 5,000 samples each per epoch.

The smoothed log-likelihood function is evaluated using kernel density estimation with $N_G=500$ generated samples from normalizing flows:
\begin{equation}
    (\hat{f}_\perp * K_h) (\vec{w}_\perp; \theta) \rightarrow \frac{1}{N_G} \sum_{g=1}^{N_G} K_h \left[ \vec{w}_\perp - \vec{w}_\perp^{(g)}(\theta) \right],
\end{equation}
where $\vec{w}_\perp^{(g)}(\theta)$ represents the generated and projected sample that is differentiable with respect to neural network parameters $\theta$.
$K_h$ is the Gaussian kernel with bandwidth $h$.
For $D$-dimensional density estimation, we use the rule-of-thumb bandwidth $h=N^{-\frac{1}{4+D}}$ for standardized data.
We set the generated dataset size equal to the training/validation dataset size to match the distribution tail support, as larger datasets extend to longer tails.

The complete loss function $\mathcal{L}(\theta)$ is then:
\begin{equation}
    - \frac{1}{N N_K} \sum_{a=1}^N \sum_{b=1}^{N_K}
    \log \left[ 
        \frac{1}{N_G} \sum_{g=1}^{N_G} K_h \left[ \vec{w}_\perp^{(a)} + \vec{\epsilon}^{(b)} - \vec{w}_\perp^{(g)}(\theta) \right]
    \right].
\end{equation}
We minimize this loss function using the Adam optimizer \cite{kingma2014adam} with learning rate 0.001.
For other parameters, we use the default parameters in \texttt{PyTorch}.
The training process is repeated 20 times with different random seeds, and we report ensemble-averaged predictions to reduce random fluctuations.

\begin{figure*}[!t]
    \centering
    \includegraphics[width=0.245\linewidth]{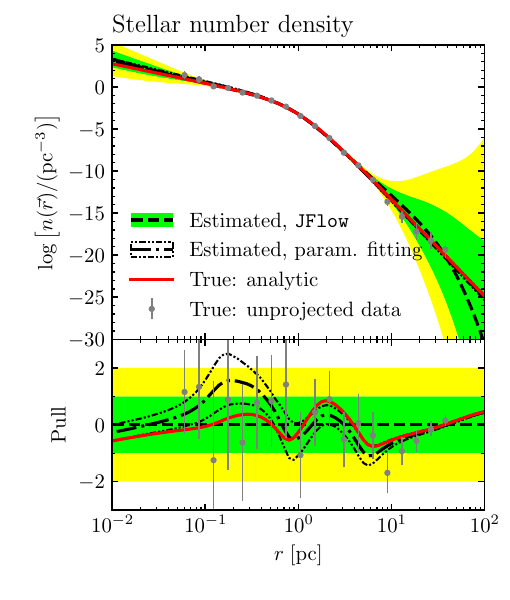}
    \includegraphics[width=0.245\linewidth]{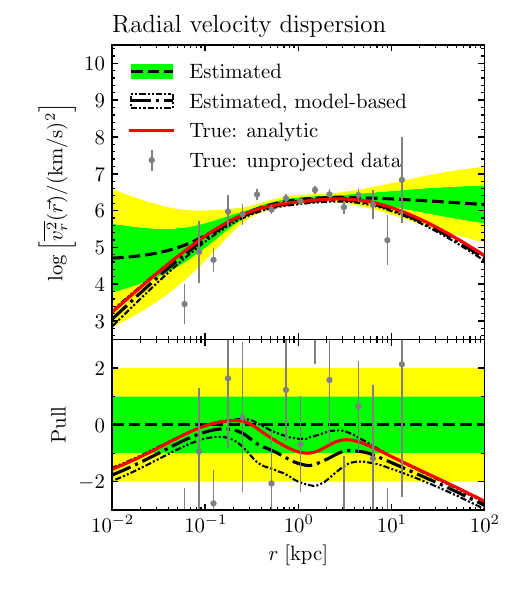}
    \includegraphics[width=0.245\linewidth]{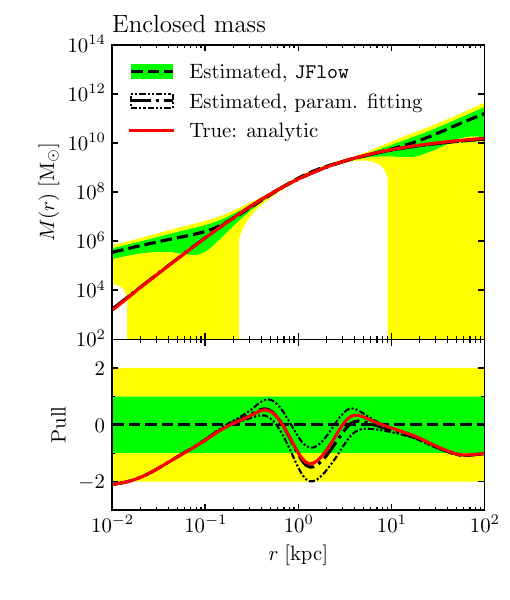}
    \includegraphics[width=0.245\linewidth]{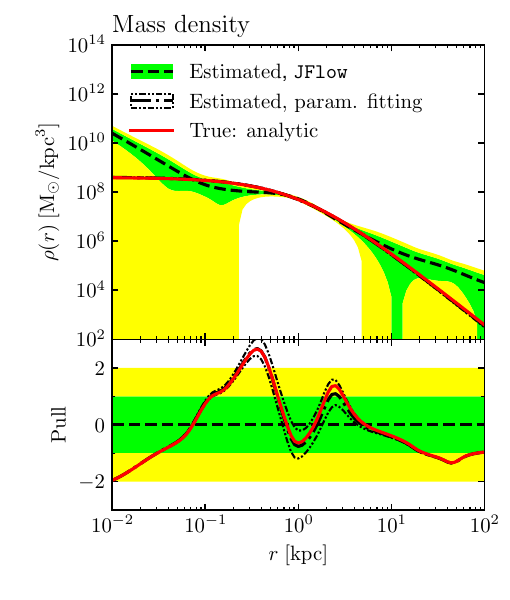}
    \caption{
        Maximum a posteriori estimation results for dSph parameters of \texttt{NonplumCoreIso} dataset with the correct velocity anisotropy function $\beta=0$, using the parametric model described in \eqref{eqn:model:stellar} and \eqref{eqn:model:dark_matter}.
        Black dot-dashed lines are the best-fit results, and black dot-dot-dashed bands are their $1\sigma$ credible intervals.
        The results are overlaid on the copy of \figref{fig:jflow_res}.
    }
    \label{fig:jflow_res:map}
\end{figure*}

\underline{Stellar number density training-specific details:}
For constructing the likelihood function, we consider two projected input features $\vec{w}_\perp$: the projected galactocentric distance $|\vec{r}_\perp|$ and its logarithm $\log_{10} |\vec{r}_\perp|$.
We simultaneously minimize two loss functions: one using $\vec{w}_\perp = |\vec{r}_\perp|$ and the other using $\vec{w}_\perp = \log_{10} |\vec{r}_\perp|$. 
While the two loss functions should converge to the same optimal solution in the asymptotic limit, we find that simultaneous minimization mildly improves regression quality at small length scales when working with limited samples.

The training of the cusp parameter $c$ in \eqref{eqn:trans_pow} is separated from the neural network training to provide additional regularization, as this parameter determines the stellar density behavior at the galactic center where we have a limited number of samples. 
We loop the following process 20 times to train the cusp parameter consistently.
\begin{enumerate}
\item
Initialize the cusp parameter $c$ by randomly sampling the cusp scaling exponent $-\frac{3c}{c+1}$ from the uniform distribution on $(-1,0]$.
\item
Fix the cusp parameter $c$ and train only the network parameters; $c$ will be updated later after finishing the network parameter training. The transition scale $r_s$ in \eqref{eqn:cnf_core} is set to the 1D rule-of-thumb bandwidth for the standardized dataset, $N^{-\frac{1}{5}}$.
\item
After training the network parameters, evaluate the cusp exponent $d \log n(r)/d \log r$ at 300 random locations sampled from a 3D Gaussian distribution centered at the galactic center, with standard deviation $N^{-\frac{1}{5}}$ for each axis.
\item
Use the mean cusp exponent $-\bar{\gamma}=E\left[ d \log n(r)/d \log r\right]$ as the next cusp scaling exponent, i.e., re-initialize $c= \frac{\bar{\gamma}}{3-\bar{\gamma}}$ for the next training iteration. Repeat steps 2-4.
\end{enumerate}
Each iteration with fixed $c$ takes approximately 300-400 seconds on an NVIDIA L40S GPU with GPU utilization about 25\% and TF32 precision, making the total stellar number density training time a few hours.

\underline{Radial velocity dispersion training-specific details:}

The training setup for radial velocity dispersion is similar to that for the stellar number density. 
We consider two sets of projected input features $\vec{w}_\perp$ for constructing two loss functions: one using $(|\vec{r}_\perp|, v_z)$ and the other $(\log{10} |\vec{r}_\perp|, v_z)$, where $v_z$ is the line-of-sight velocity.
The remaining implementation details follow the stellar number density training procedure, but without the extra looping process since the velocity dispersion model has no parameters requiring special treatment of low-statistics regions.
The training takes approximately one hour on an NVIDIA L40S GPU, bringing the total \texttt{JFlow} training time to less than 6 hours per one base learner in ensemble averaging.

\section{Comparing \texttt{JFlow} with Conventional Parametric Inference Methods}
\label{app:map}

As the number of stars in dSph stellar catalogs is typically small, conventional analyses fit parameters of stellar and dark matter distribution models to infer dSph properties through Jeans analysis \cite{Hayashi:2016kcy, Ichikawa:2016nbi, Ichikawa:2017rph, Horigome:2020kyj}.
In this appendix, we compare our \texttt{JFlow} method with a conventional parametric inference method.
We use the stellar number density model in \eqref{eqn:model:stellar} and the dark matter density model in \eqref{eqn:model:dark_matter} as our parametric model setup.
For stellar number density, we apply the Abel transformation to evaluate the projected likelihood, as we only observe positions perpendicular to the line-of-sight direction.
For radial velocity dispersion, we solve the spherical Jeans equation to obtain velocity dispersions for this parametric model, then apply the Abel transformation to model the line-of-sight velocity dispersion \cite{Ullio_2016}. 
Since we are modeling velocity variance, we fit the data using a Gaussian likelihood with the line-of-sight velocity dispersion.

The model contains nine parameters, and we perform maximum a posteriori estimation to find their best-fit values and uncertainties.
We set uniform priors for all parameters as listed in \tableref{tab:prior}.
We then sample the posterior probability using importance sampling with a Gaussian proposal distribution centered at the true values.
The width of each Gaussian for each axis is set to the largest parameter offset that increases the log-likelihood by 0.5 from its maximum.
We start with 600,000 initial trial samples, and the sample size is reduced to 640 after unweighting.
While the posterior sampling process takes approximately one day, this could be accelerated using Markov Chain Monte Carlo methods and more sophisticated numerical integration techniques rather than direct integration.

\figref{fig:jflow_res:map} shows the maximum a posteriori estimation results in dot-dashed lines.
The parametric inference predicts narrower credible intervals than the confidence intervals of \texttt{JFlow}.
This is expected from the bias-variance tradeoff: the parametric model considers only specific functional forms, leading to smaller uncertainties than more expressive models like \texttt{JFlow}.
The underestimated systematic uncertainties in profile shapes are a general feature when comparing parametric models with free-form approaches, as also observed in Milky Way analyses \cite{Lim:2023lss}.

In the fitting above, we use a function model that includes the true solution, resulting in good agreement with the true values.
However, if the true solution lies outside the assumed functional form, parametric inference will be biased.
\texttt{JFlow} is a general free-form model for spherical Jeans analysis, providing robust estimation when the true distribution is unknown.
This flexibility makes our method particularly valuable when analyzing various galaxy types with unknown dark matter distributions, enabling tests of different physical scenarios.

\end{document}